\documentclass[letterpaper,journal]{IEEEtran}
\usepackage{amsmath,amsfonts,amssymb}
\usepackage{algorithmic}
\usepackage{algorithm}
\usepackage{array}
\usepackage[caption=false,font=normalsize,labelfont=sf,textfont=sf]{subfig}
\usepackage{textcomp}
\usepackage{stfloats}
\usepackage{url}
\usepackage{verbatim}
\usepackage{graphicx}
\usepackage{cite}
\usepackage{balance}
\usepackage[
colorlinks=true,
linkcolor=red,
citecolor=blue,
urlcolor=blue
]{hyperref}
\usepackage[font=small,skip=2pt]{caption}

\usepackage{enumitem}
\setlist{nosep}
\begin{document}

\title{Fluid Antenna Channel Reconstruction \\with Non-Negative Least Square Detector}

\author{Yang Chen, Jian Dang, Zaichen Zhang
	\thanks{Yang Chen is with the National Mobile Communications Research Laboratory, Frontiers Science Center for Mobile Information Communication and Security, Southeast University, Nanjing 210096, China (e-mail: q2460846608@163.com).}
	
	\thanks{Zaichen Zhang is with the National Mobile Communications Research Laboratory, Frontiers Science Center for Mobile Information Communication and Security, Southeast University, Nanjing 210096, China. Zaichen Zhang is also with Purple Mountain Laboratories, Nanjing 211111, China. (e-mail: zczhang@seu.edu.cn).}
	
	\thanks{Jian Dang is with the National Mobile Communications Research Laboratory, Frontiers Science Center for Mobile Information Communication and Security, Southeast University, Nanjing 210096, China, also with the Key Laboratory of Intelligent Support Technology for Complex Environments, Ministry of Education, Nanjing University of Information Science and Technology, Nanjing 210044, China, and also with Purple Mountain Laboratories, Nanjing 211111, China. (e-mail: dangjian@seu.edu.cn).}
	
}

\maketitle

\begin{abstract}
	Reconstructing the full-aperture channel of a fluid antenna system (FAS)
	from a few activated ports requires inferring unobserved responses from
	limited spatial samples. This paper proposes a covariance-domain FAS
	channel-reconstruction method based on nonnegative least squares (NNLS).
	A nonuniform port arrangement constructs a difference coarray with an
	enlarged aperture. Vectorizing the sample covariance maps the physical
	array to a virtual steering dictionary, and the resulting complex
	covariance-fitting model is recast as a strictly equivalent real-valued
	NNLS problem. A projected-gradient iteration initialized at zero estimates
	the nonnegative angular-occurrence coefficients, and the $L_s$ strongest
	well-separated peaks yield the angles of arrival (AoAs). Simulations show
	that the proposed FAS-NNLS estimator exploits the enlarged virtual aperture
	and achieves lower AoA mean squared errors than uniform linear array (ULA)
	multiple signal classification (ULA-MUSIC) and ULA estimation of signal
	parameters via rotational invariance techniques (ULA-ESPRIT). Its
	non-line-of-sight (NLoS) path accuracy matches FAS covariance matching
	pursuit, while its line-of-sight (LoS) estimation is significantly improved.
\end{abstract}


\section{Introduction}
\label{sec:introduction}

Uniform linear arrays (ULAs) with half-wavelength spacing are a canonical
architecture for spatial signal processing. Their Vandermonde steering
structure underpins mature direction-of-arrival estimators, including
multiple signal classification (MUSIC) and estimation of signal parameters
via rotational invariance techniques (ESPRIT)
\cite{schmidt1986music,roy1989esprit}. However, fixed element positions
predetermine both the aperture and the sampling pattern for a given element
count, limiting the spatial degrees of freedom of compact devices.

Fluid antenna systems (FASs)\cite{fas-twc-21,kit_electronic,ZZT_FiniteBlocklengthFAS_2026,ZZT_FiniteBlocklengthCorrelation_2026} relax this constraint by switching a limited
number of radio-frequency chains among a dense set of candidate ports in a
prescribed region \cite{wong2023preliminaries,ZZT_FiniteAperturePlanar_2026,ZZT_FiniteApertureDesign_2026,FAA_XJY}. The activated port
locations thus become additional design variables: nonuniform patterns
reshape the aperture and its difference set without increasing the active
port count \cite{ZZT_SlowFluidURA_2025,11586648,11155198}, allowing a small number of observations to
preserve useful propagation information.

Representative FAS reconstruction methods include finite-path sparse
estimation \cite{xu2024l3scr,11556494,ZZT_DualSideFAS_2026}, electromagnetic-compliant Nyquist and
maximum-likelihood reconstruction \cite{new2025oversampling,ML_WCNC}, and the
successive Bayesian reconstructor \cite{ZZT_JointActivityChannel_2026,ZZT_LearnedAMP_2026,ZZT_GeometryReconstruction_2026}. These approaches
mainly operate in the physical-port domain, leaving the virtual aperture
gained from nonuniform FAS activation underexploited. Vectorizing the covariance of a nonuniform array maps each port pair to a
difference lag, constructing a virtual difference coarray. Under a
finite-path model with independent path coefficients, the covariance is a
nonnegative combination of virtual steering responses, making NNLS fitting
a natural convex means of recovering the angular support over the enlarged
virtual aperture.

The main contributions of this paper are as follows.
\begin{itemize}
	\item We link nonuniform FAS port selection to covariance-domain
	reconstruction by vectorizing the received covariance into a virtual
	difference-coarray model with an explicit noise basis component.
	\item We formulate angle-of-arrival (AoA) support recovery as an
	equivalent real-valued NNLS problem solved by projected-gradient
	iteration from the zero vector, with the strongest separated peaks
	determining the detected AoAs.
	\item We complete parametric channel reconstruction via least-squares
	path-gain estimation and Rice-factor-based line-of-sight (LoS)
	identification. Comparisons with ULA-MUSIC, ULA-ESPRIT, and FAS
	covariance matching pursuit confirm that FAS-NNLS translates the
	virtual aperture gain into improved AoA accuracy.
\end{itemize}

\section{Fluid Antenna System Model}
\label{sec:fas_system_model}

\subsection{Fluid Antenna Channel Model}
\label{subsec:fas_channel_model}

\begin{figure}[!t]
	\centering
	\includegraphics[width=\columnwidth]{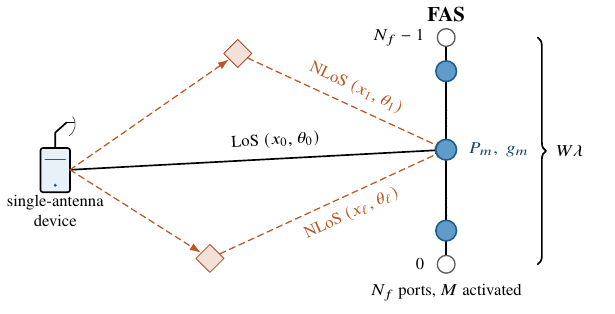}
	\caption{Uplink FAS channel model (Filled and hollow circles denote activated and
	candidate ports).}
	\label{fig:fas_channel_model}
\end{figure}

Consider an uplink transmission from a single-antenna device to a fluid
antenna base station with $N_f$ configurable ports, of which only $M$ ports
are activated during channel training. The corresponding propagation
geometry and port-activation model are illustrated in
Fig.~\ref{fig:fas_channel_model}.

Let
$\mathbf{g}=\left[g_1,g_2,\ldots,g_M\right]^{\mathrm{T}}
\in\mathbb{C}^{M\times 1}$ collect the channel coefficients over the
activated ports. Under planar-wave propagation, its $m$-th element is
written as
\begin{equation}
	\label{eq:1}
	g_{m}=\underbrace{x_{0}e^{-j\frac{2\pi P_m W}{N_f-1}\sin\theta_{0}}}_{\text{LoS component}}+\underbrace{\sum_{l=1}^{L_s-1}x_{l}e^{-j\frac{2\pi P_mW}{N_f-1}\sin\theta_{l}}}_{\text{non-LoS component}},
\end{equation}
where the model consists of one specular component (LoS) and finite $L_s-1$ scatterer components (non-LoS), $x_0=\sqrt{\frac{K}{K+1}}e^{j\alpha}$ denotes the path strength with Rice factor $K$\cite{wong2022extra}, $x_l,l\in [1:L_s]$ is the strength of the $l$-th scattered path with constraint $\sum_{l=1}^{L_s}|x_l|^2=\frac{1}{K+1}$, $P_m\in[0,N_f-1]$ is the $m$-th activated array location index, $\theta_{0},\theta_l\in[-90^{\circ},90^{\circ}]$ respectively denote the azimuth angles of arrivals (AoAs) and $W=\frac{M-1}{2}$ indicates the antenna length normalized by wavelength. Expanding the channel vector into matrix form and we can decompose the $\mathbf{g}$ into the product of matrix $\mathbf{W}=\left[\mathbf{w}_0,\mathbf{w}_1,\ldots,\mathbf{w}_{L_s-1}\right]\in \mathbb{C}^{M\times L_s}$ and path strength vector $\mathbf{x}\in \mathbb{C}^{L_s\times 1}$, i.e., $\mathbf{g}=\mathbf{W}\mathbf{x}$:
\begin{equation}
	\label{eq:2}
	\begin{aligned}
		&\mathbf{g}=\begin{bmatrix}
			g_{1}\\
			g_{2}\\
			\vdots\\
			g_{M}
		\end{bmatrix}
		=
		\begin{bmatrix}
			\sum_{l=0}^{L_s}x_{l} e^{-j\frac{2\pi P_1 W}{N_f-1}\sin\theta_{l}}\\
			\sum_{l=0}^{L_s}x_{l} e^{-j\frac{2\pi P_2 W}{N_f-1}\sin\theta_{l}}\\
			\vdots\\
			\sum_{l=0}^{L_s}x_{l} e^{-j\frac{2\pi P_M W}{N_f-1}\sin\theta_{l}}
		\end{bmatrix}= \\
		&\underbrace{\begin{bmatrix}
				e^{-j\frac{2\pi P_1 W}{N_f-1}\sin\theta_{0}} &   \ldots & e^{-j\frac{2\pi P_1 W}{N_f-1}\sin\theta_{L_s}}\\
				e^{-j\frac{2\pi P_2 W}{N_f-1}\sin\theta_{0}} & \ldots & e^{-j\frac{2\pi P_2W}{N_f-1}\sin\theta_{L_s}}\\
				\vdots &  \ddots  & \vdots\\
				e^{-j\frac{2\pi P_M W}{N_f-1}\sin\theta_{0}}  & \ldots & e^{-j\frac{2\pi P_M W}{N_f-1}\sin\theta_{L_s}}
			\end{bmatrix}
			\begin{bmatrix}
				x_{0}\\
				x_{1}\\
				\vdots\\
				x_{L_s}
		\end{bmatrix}}_{\mathbf{W}\mathbf{x}},
	\end{aligned}
\end{equation}
where $P_m$ is determined by the FAS port-selection pattern, the columns
of $\mathbf{W}$ are the steering vectors of the propagation paths, and
the path-gain vector satisfies
$\|\mathbf{x}\|_2^2=\sum_{l=0}^{L_s-1}|x_l|^2=1$. Once the AoAs are
available, $\mathbf{x}$ can subsequently be recovered by LS. Moreover,
because every steering vector depends on the activated indices
$\{P_m\}_{m=1}^{M}$, the port-selection pattern directly determines the
physical aperture and the resulting angular identifiability. This
observation motivates the port-selection strategies introduced next.

\subsection{Fluid Antenna Port Selection}
\label{subsec:fas_port_selection}
To exploit the spatial flexibility of the FAS, three representative
port-selection strategies are considered under the same aperture
constraint of $W$ wavelengths. Define
$\mathcal{P}=\{P_m:m=1,\ldots,M\}$, with
$\mathcal{P}\subseteq\{0,\ldots,N_f-1\}$ and
$|\mathcal{P}|=M$, as the set of activated port indices.
\subsubsection{Optimal Selection}
\label{subsubsec:optimal_selection}
If the channel reconstruction can be conducted perfectly or the channel state information between all configurable ports and device are estimated, the array responses at all ports can be acquired, which though requires too much overhead in practice. Nevertheless, the optimal selection in fluid antenna often refers to activating $M$ array elements whose channel responses are the largest among all candidates: 
\begin{equation}
	\label{eq:3}
	\mathcal{P}=\mathop{\arg\max}_\text{largest $M$ ones}\left \{|\hat{g}_{1}|,|\hat{g}_{2}|,...,|\hat{g}_{N_f}|\right \}.
\end{equation}
\subsubsection{Uniform Selection}
\label{subsubsec:uniform_selection}
Since the steering vectors are determined by AoAs and path strength, one can restore the overlapped channel vector via estimating only a portion of the ports to restore the array response at every candidates. One feasible way is to estimate $M$ ports with $\Delta$ index gap. If the ports are activated every $\Delta$ ports from the 1st one, the ports location equals:
\begin{equation}
	\label{eq:4}
	\mathcal{P}=\left \{(m-1)\Delta, m\in [1,M]\right \}, \Delta\le \left \lfloor \frac{N_f-1}{M-1} \right \rfloor 
\end{equation}
where the gap $\Delta$ is an integer to be determined to satisfy certain goals.
\subsubsection{Non-Uniform Selection}
\label{subsubsec:nonuniform_selection}
While many state-of-the-arts focusing on the aforementioned optimal and uniform ports permutation, the non-uniform selection is particularly relevant to this work. In the spirit of  
difference co-array (DCA) design \cite{11155198}, favorable non-uniform selection aims to reduce the redundancy and increase the DoFs in the weight function. The weight function is defined as $w(x)$ where $x=(P_m-P_n),m,n\in [1,M]$ denote how many pairs of array location $P_m,P_n$ with difference of $x$. Let $\mathcal{P}_{w}$ denote all potential possibilities location difference, i.e., $P_m-P_n$. For any $x\in \mathcal{P}_{w}$, weight function $w(x)$ manifests the following features: 
\begin{subequations}
	\label{f}
	\begin{align}
		&w(0)=M
		\label{eq:5a},\\
		&\forall x\in \mathcal{P}_{w}\setminus \{0\}, 1\le w(x) \le M-1
		\label{eq:5b},\\
		&\forall x\in \mathcal{P}_{w}, w(x)=w(-x)
		\label{eq:5c},\\
		&\sum_{x\in \mathcal{P}_{w},x\neq 0}w(x)=M(M-1)
		\label{eq:5d}.
	\end{align}
\end{subequations}

By the \eqref{eq:5d}, one can observe a maximum DoF of $\mathcal{P}_{w}$ in the order of $M(M-1)+1$. When $w(x)>1,\forall x\in \mathcal{P}_{w}\setminus \{0\}$, the DoF decreases. Therefore, it's advisable to maintain $w(x)\rightarrow 1,\forall x\in \mathcal{P}_{w}\setminus \{0\}$ as many as possible. In fact, for small- and medium-scale fluid antenna arrays, optimal MRA-based non-uniform port activation patterns for several given values of $M$ have been reported in \cite{11155198}. Such a nonuniform activation pattern supplies the enlarged difference-coarray aperture used by the covariance-domain reconstruction model developed next.
\section{Channel Reconstruction Problem}
\label{sec:channel_reconstruction_problem}
Building on the channel and port-selection models in
Section~\ref{sec:fas_system_model}, this section formulates the uplink
training model \cite{ZZT_CRBActivityDetection_2026} and converts the received-signal covariance into a
virtual-array observation with an enlarged effective aperture.

Let $\mathbf{s}\in \mathbb{C}^{1\times L}$ denote the pilot signal vector with $L$ snapshots/channel uses following distribution $\mathcal{CN}(0,\mathbf{I})$ with zero mean and unit covariance matrix. The received signal $\mathbf{Y}\in \mathbb{C}^{M\times L}$ can be written as:
\begin{equation}
	\label{eq:6}
	\mathbf{Y} = \sqrt{\eta\beta}\mathbf{g}\mathbf{s}+\mathbf{N},
\end{equation}
where $\eta$ is the averaged signal power, $\beta=d^{-\gamma}10^{\xi/10}$ is the large scale lognormal fading factor with $d$ the distance between device and relay in meters, $\gamma$ the path loss exponent and $\xi$ a random variable following distribution of $\mathcal{N}(1,\delta)$ and $\mathbf{N}$ is the additive white Gaussian noise following $\mathcal{CN}(0,\sigma^2\mathbf{I})$ with zero mean and variance $\sigma^2$. Therefore, by assuming the independence among signal and noise components, the noisy covariance matrix $\widetilde{\mathbf{R}}_{YY}$ of the received signal can be approximated as:
\begin{equation}
	\label{eq:7}
	\begin{aligned}
		\widetilde{\mathbf{R}}_{YY}&=\mathbb{E}\left\{\mathbf{Y}\mathbf{Y}^{\mathrm{H}} \right\},\\
		&=\mathbb{E}\left\{\left(\sqrt{\eta\beta}\mathbf{g}\mathbf{s}+\mathbf{N}\right)\left(\sqrt{\eta\beta}\mathbf{g}\mathbf{s}+\mathbf{N}\right)^{\mathrm{H}}\right\},\\
		&=\mathbb{E}\left\{\eta\beta\mathbf{gs}\mathbf{s}^{\mathrm{H}}\mathbf{g}^{\mathrm{H}}+L\sigma^2\mathbf{I}_M\right\},\\
		&=L\eta\beta\mathbb{E}\left\{\mathbf{g}\mathbf{g}^{\mathrm{H}}\right\}+L\sigma^2\mathbf{I}_M.
	\end{aligned}
\end{equation}
Substitute $\|\mathbf{x}\|_2^2=\sum_{l=0}^{L_s-1}x_l^2=1$ and $\mathbf{g}=\mathbf{W}\mathbf{x}$ into \eqref{eq:4} and consider the randomized transmission paths, i.e., $x_0,x_1,\ldots,x_{L_s-1}$ are deemed as arbitrary and independent random variables:
\begin{equation}
	\label{eq:8}
	\begin{aligned}
		\widetilde{\mathbf{R}}_{YY}&= L\eta\beta\mathbb{E}\left\{\mathbf{W}\mathbf{W}^{\mathrm{H}}\right\}+L\sigma^2\mathbf{I}_M,\\
		&=L\eta\beta\sum_{l=0}^{L_s-1}\mathbf{w}_l\mathbf{w}^{\mathrm{H}}_l+L\sigma^2\mathbf{I}_M,\\
	\end{aligned}
\end{equation}
where $\mathbf{w}_l$ is the $l$-th steering vector in \eqref{eq:2} and determined by AoA $\theta_l$ and activated array location $P_m$.

Let $\mathbf{R}_{YY}=\sum_{l=0}^{L_s-1}\mathbf{w}_l\mathbf{w}^{\mathrm{H}}_l\in \mathbb{C}^{M\times M}$ denote the covariance matrix of the signal components, i.e., $\frac{1}{L\eta\beta}\widetilde{\mathbf{R}}_{YY}=\mathbf{R}_{YY}+\frac{\sigma^2}{\eta\beta}\mathbf{I}_M$. Focusing on the element at $m$-th row and $n$-th column in $\mathbf{R}_{YY}$, $m,n\in[1,M]$:
\begin{equation}
	\label{eq:9}
	\left[\mathbf{R}_{YY}\right]_{m,n}=\sum_{l=0}^{L_s}e^{-j\frac{2\pi (P_m-P_n) W}{N_f-1}\sin\theta_{l}},
\end{equation}
where $P_m-P_n$ is the array location difference. Notably, by vectorization, the original sensing signal model in \eqref{eq:7} can be converted into a virtual array model $\mathbf{r}\in \mathbb{C}^{M^2\times 1}$ with DCA elements in $\mathcal{P}_{w}$:
\begin{equation}
	\label{eq:10}
	\begin{aligned}
		\mathbf{r}&=\mathrm{vec}\left(\frac{1}{L\eta\beta}\widetilde{\mathbf{R}}_{YY}\right)
		=\mathrm{vec}\left(\mathbf{R}_{YY}+\frac{\sigma^2}{\eta\beta}\mathbf{I}_M \right),\\
		&=\mathrm{vec}\left(\sum_{l=0}^{L_s}\mathbf{w}_l\mathbf{w}^{\mathrm{H}}_l+\frac{\sigma^2}{\eta\beta}\mathbf{I}_M\right),\\
		&=\sum_{l=0}^{L_s-1}\mathbf{w}_l^*\otimes \mathbf{w}_l+\frac{\sigma^2}{\eta\beta}\mathbf{I}_{M^2},
	\end{aligned}
\end{equation}
where $\otimes$ is Kronecker product and $\mathbf{I}_{M^2}=\left[\mathbf{e}_1^{\mathrm{T}},\mathbf{e}_2^{\mathrm{T}},\ldots, \mathbf{e}_M^{\mathrm{T}}\right]$ and column vector $\mathbf{e}_i,i\in [1:M]$ are indicator vector with element at $i$-th location equal to 1 and others all zeros. Hereby, the virtual array signal in \eqref{eq:10} has wider receiving aperture than the original model in \eqref{eq:8}, i.e., $M^2> M$. 

Specifically, there are only finite scatterers in \eqref{eq:10}, i.e., $L_s$ different $\mathbf{w}_l^*\otimes \mathbf{w}_l$ to be detected. Thereby, we can treat the AoA detection as a sparse recovery problem. Let $\mathbf{A}=\left[\mathbf{W}^*\odot \mathbf{W},\frac{\sigma^2}{\eta\beta}\mathbf{I}_{M^2}\right]\in \mathbb{C}^{M^2\times N+1}$ be a codebook where $\mathbf{W}^*\odot \mathbf{W}=\left[\mathbf{w}_1^*\otimes\mathbf{w}_1, \mathbf{w}_2^*\otimes\mathbf{w}_2,\ldots,\mathbf{w}_N^*\otimes\mathbf{w}_N\right]$, $\odot$ is the Khatri-Rao product, $\mathbf{I}_{M^2}$ denotes the vectorized noise basis and $N\gg L_s$ is the number of AoA samples. Thereupon, the AoA detection can be written into a compact form:
\begin{equation}
	\label{eq:11}
	\mathbf{r}
	=
	\mathbf{A}
	\underbrace{
	\begin{bmatrix}
		\mathbf{b}\\
		1
	\end{bmatrix}
	}_{\mathbf{e}},
\end{equation}
where $\mathbf{b}\in\mathbb{R}^{N\times1}$ denotes the sparse angular-domain coefficient vector, while the last entry of $\mathbf{e}\in\mathbb{R}^{(N+1)\times1}$ corresponds to the known noise-basis component. In the ideal on-grid case, $b_n=1$ if the $n$-th candidate angle corresponds to an existing propagation path, and $b_n=0$ otherwise. Consequently, $\|\mathbf{b}\|_0=L_s$, whereas the augmented vector $\mathbf{e}$ is $(L_s+1)$-sparse due to the additional noise-basis component. Therefore, the recovery of $\mathbf{b}$ can be formulated as a nonnegative sparse recovery problem, as detailed in the sequel.

\section{Channel Reconstruction Detector}
\label{sec:channel_reconstruction_detector}

\subsection{Benchmark Algorithms}
\label{subsec:benchmark_algorithms}

Three benchmarks are adopted: ULA-MUSIC, ULA-ESPRIT, and FAS covariance
matching pursuit (MP). The first two use a half-wavelength ULA; the third
shares the FAS covariance dictionary with the proposed method.

For the ULA benchmarks, the steering vector and sample covariance are
\begin{equation}
	\label{eq:benchmark_ula_model}
	\mathbf{v}(\theta)
	=
	\begin{bmatrix}
		1, \cdots,
		e^{-j\pi(M-1)\sin\theta}
	\end{bmatrix}^{\mathrm{T}},~
	\widehat{\mathbf{R}}_{\mathrm{U}}
	=
	\frac{1}{L}\mathbf{Y}_{\mathrm{U}}\mathbf{Y}_{\mathrm{U}}^{\mathrm{H}},
\end{equation}
where $\mathbf{Y}_{\mathrm{U}}$ follows \eqref{eq:6} with
$\mathbf{v}(\theta)$ replacing the FAS steering vector. The scalar
factor $\eta\beta$ does not affect the estimated eigenspaces.

\subsubsection{ULA-MUSIC}
\label{subsubsec:ula_music}
Let $\widehat{\mathbf{U}}_{\mathrm{n}}$ contain the eigenvectors of
$\widehat{\mathbf{R}}_{\mathrm{U}}$ associated with its $M-L_s$ smallest
eigenvalues. The MUSIC spectrum is \cite{schmidt1986music}
\begin{equation}
	\label{eq:music_spectrum}
	P_{\mathrm{MUSIC}}(\theta)
	=
	\frac{1}{
		\mathbf{v}^{\mathrm{H}}(\theta)
		\widehat{\mathbf{U}}_{\mathrm{n}}
		\widehat{\mathbf{U}}_{\mathrm{n}}^{\mathrm{H}}
		\mathbf{v}(\theta)}.
\end{equation}
The $L_s$ strongest separated maxima of \eqref{eq:music_spectrum} give
the AoA estimates.

\subsubsection{ULA-ESPRIT}
\label{subsubsec:ula_esprit}
Let $\widehat{\mathbf{U}}_{\mathrm{s}}$ hold the $L_s$ dominant
eigenvectors of $\widehat{\mathbf{R}}_{\mathrm{U}}$, and define
$\widehat{\mathbf{U}}_1 = \widehat{\mathbf{U}}_{\mathrm{s}}(1\!:\!M-1,:)$
and
$\widehat{\mathbf{U}}_2 = \widehat{\mathbf{U}}_{\mathrm{s}}(2\!:\!M,:)$.
The least-squares ESPRIT estimator, assuming
$\operatorname{rank}(\widehat{\mathbf{U}}_1)=L_s$, is
\cite{roy1989esprit}
\begin{equation}
	\label{eq:esprit_benchmark}
	\begin{aligned}
		\widehat{\boldsymbol{\Psi}}
		&=
		\!\left(
		\widehat{\mathbf{U}}_1^{\mathrm{H}}
		\widehat{\mathbf{U}}_1
		\right)^{-1}
		\widehat{\mathbf{U}}_1^{\mathrm{H}}
		\widehat{\mathbf{U}}_2,
		\\
		\widehat{\theta}^{\mathrm{ESPRIT}}_l
		&=
		\arcsin\!\left(
		-\frac{\angle\widehat{\zeta}_l}{\pi}
		\right),~ l=1,\ldots,L_s,
	\end{aligned}
\end{equation}
where $\widehat{\zeta}_l$ is the $l$-th eigenvalue of
$\widehat{\boldsymbol{\Psi}}$. ESPRIT exploits ULA shift invariance and
requires no grid search.

\subsubsection{FAS Covariance Matching Pursuit}
\label{subsubsec:fas_covariance_mp}
Let $\widehat{\mathbf{r}}$ be the sample FAS covariance vector in
\eqref{eq:10}, and let $\boldsymbol{\phi}_n$ be the $n$-th column of
$\boldsymbol{\Phi}=\mathbf{W}^{*}\odot\mathbf{W}$ in \eqref{eq:11}.
Starting from $\mathcal{S}^{(0)}=\varnothing$ and
$\mathbf{q}^{(0)}=\widehat{\mathbf{r}}$, each iteration performs
\cite{9449971}
\begin{equation}
	\label{eq:fas_mp_iteration}
	\begin{aligned}
		\widehat{n}_k
		&=
		\underset{n\notin\mathcal{S}^{(k-1)}}{\arg\max}
		\left|\boldsymbol{\phi}_n^{\mathrm{H}}\mathbf{q}^{(k-1)}\right|,
		\\
		\mathcal{S}^{(k)}
		&=\mathcal{S}^{(k-1)}\cup\{\widehat{n}_k\},
		\\
		\widehat{\mathbf{c}}_{\mathcal{S}^{(k)}}
		&=
		\!\left(
		\boldsymbol{\Phi}_{\mathcal{S}^{(k)}}^{\mathrm{H}}
		\boldsymbol{\Phi}_{\mathcal{S}^{(k)}}
		\right)^{-1}
		\boldsymbol{\Phi}_{\mathcal{S}^{(k)}}^{\mathrm{H}}
		\widehat{\mathbf{r}},
		\\
		\mathbf{q}^{(k)}
		&=\widehat{\mathbf{r}}-
		\boldsymbol{\Phi}_{\mathcal{S}^{(k)}}
		\widehat{\mathbf{c}}_{\mathcal{S}^{(k)}}.
	\end{aligned}
\end{equation}
After $L_s$ iterations, the indices in $\mathcal{S}^{(L_s)}$ give the
detected AoAs via greedy selection with joint least-squares coefficient
refitting.

All three benchmarks differ from the proposed method only in the AoA
estimation stage; their estimates are subsequently processed by the common
path-gain estimation and LoS identification procedure in
Subsection~\ref{subsec:ls_los}.

\subsection{NNLS-Based AoA Estimation}
\label{subsec:nnls_aoa_estimation}

To explicitly formulate the sparse recovery problem in \eqref{eq:11}, 
let $\{\theta_n\}_{n=1}^{N}\subseteq[-\pi/2,\pi/2]$ denote the 
predefined AoA candidates. The physical-array steering vector associated
with the $n$-th candidate angle is given by
\begin{equation}
	\label{eq:12}
	\mathbf{w}(\theta_n)
	=
	\begin{bmatrix}
		e^{-j\frac{2\pi P_1W}{N_f-1}\sin\theta_n}\\
		e^{-j\frac{2\pi P_2W}{N_f-1}\sin\theta_n}\\
		\vdots\\
		e^{-j\frac{2\pi P_MW}{N_f-1}\sin\theta_n}
	\end{bmatrix}
	\in\mathbb{C}^{M\times 1}.
\end{equation}
Accordingly, its virtual-array response is
\begin{equation}
	\label{eq:13}
	\mathbf{a}(\theta_n)
	=
	\mathbf{w}^{*}(\theta_n)
	\otimes
	\mathbf{w}(\theta_n)
	\in\mathbb{C}^{M^2\times 1}.
\end{equation}
Collecting the virtual-array responses of all candidate angles yields
the angular dictionary
\begin{equation}
	\label{eq:14}
	\boldsymbol{\Phi}
	=
	\begin{bmatrix}
		\mathbf{a}(\theta_1) &
		\mathbf{a}(\theta_2) &
		\cdots &
		\mathbf{a}(\theta_N)
	\end{bmatrix}
	\in\mathbb{C}^{M^2\times N}.
\end{equation}

For notational clarity, define the vectorized noise scaled version as
\begin{equation}
	\qquad
	\mathbf{a}_{\mathrm n}
=
	\frac{\sigma^2}{\eta\beta}\mathbf{I}_{M^2}.
\end{equation}
Then, the codebook in \eqref{eq:11} can be partitioned as
\begin{equation}
	\label{eq:16}
	\mathbf{A}
	=
	\begin{bmatrix}
		\boldsymbol{\Phi} & \mathbf{a}_{\mathrm n}
	\end{bmatrix}.
\end{equation}
Since the last entry of $\mathbf{e}$ is fixed to one and the noise
variance is assumed to be known, the noise contribution can first be
removed from $\mathbf{r}$ as
\begin{equation}
	\label{eq:19}
	\mathbf{r}_{\mathrm s}
=
	\mathbf{r}-\mathbf{a}_{\mathrm n}
	=
	\boldsymbol{\Phi}\mathbf{b}.
\end{equation}
In the presence of covariance estimation errors and noise residuals,
the equality in \eqref{eq:19} is generally approximate. Therefore, the
angular coefficient vector can be estimated by fitting
$\boldsymbol{\Phi}\mathbf{b}$ to $\mathbf{r}_{\mathrm s}$.

Although $\mathbf{r}_{\mathrm s}$ and $\boldsymbol{\Phi}$ are complex
valued, $\mathbf{b}$ represents the nonnegative occurrence weights of
the candidate angles and is thus real valued. Define the following
real-valued observation vector and dictionary:
\begin{equation}
	\label{eq:20}
	\mathbf{y}
=
	\begin{bmatrix}
		\operatorname{Re}\{\mathbf{r}_{\mathrm s}\}\\
		\operatorname{Im}\{\mathbf{r}_{\mathrm s}\}
	\end{bmatrix}
	\in\mathbb{R}^{2M^2\times1},~
	\mathbf{C}
=
	\begin{bmatrix}
		\operatorname{Re}\{\boldsymbol{\Phi}\}\\
		\operatorname{Im}\{\boldsymbol{\Phi}\}
	\end{bmatrix}
	\in\mathbb{R}^{2M^2\times N}.
\end{equation}
For any $\mathbf{b}\in\mathbb{R}^{N\times1}$, the complex-valued
residual satisfies
\begin{align}
	\label{eq:21}
	\left\|
	\mathbf{r}_{\mathrm s}-\boldsymbol{\Phi}\mathbf{b}
	\right\|_2^2
	=
	\left\|\mathbf{y}-\mathbf{C}\mathbf{b}\right\|_2^2.
\end{align}
Consequently, the complex covariance-fitting problem is strictly
equivalent to the following real-valued nonnegative least-squares
(NNLS) problem \cite{zheng2023fast}:
\begin{equation}
	\label{eq:22}
	\begin{aligned}
		\underset{\mathbf{b}}{\operatorname{min}}~
		& f(\mathbf{b})
		=
		\frac{1}{2}
		\left\|\mathbf{y}-\mathbf{C}\mathbf{b}\right\|_2^2\\
		\mathrm{s.t.}~
		& \mathbf{b}\geq\mathbf{0}.
	\end{aligned}
\end{equation}
The optimization variable is $\mathbf{b}\in\mathbb{R}^{N\times1}$,
the objective function minimizes the covariance-fitting residual, and
the sole constraint enforces the nonnegativity of every angular
coefficient. Let $\widehat{\mathbf{b}}$ denote an optimal solution of
\eqref{eq:22}. The gradient of its objective function is
\begin{equation}
	\label{eq:25}
	\nabla f(\mathbf{b})
	=
	\mathbf{C}^{\mathrm T}
	\left(\mathbf{C}\mathbf{b}-\mathbf{y}\right).
\end{equation}
The Hessian matrix is
\begin{equation}
	\nabla^2 f(\mathbf{b})
	=
	\mathbf{C}^{\mathrm T}\mathbf{C}.
\end{equation}
For any $\mathbf{z}\in\mathbb{R}^{N\times1}$, its associated quadratic
form satisfies
\begin{equation}
	\mathbf{z}^{\mathrm T}
	\nabla^2 f(\mathbf{b})
	\mathbf{z}
	=
	\mathbf{z}^{\mathrm T}\mathbf{C}^{\mathrm T}\mathbf{C}\mathbf{z}
	=
	\|\mathbf{C}\mathbf{z}\|_2^2
	\geq 0.
\end{equation}
Therefore, $\nabla^2 f(\mathbf{b})$ is positive semidefinite and the
NNLS problem is convex. Writing the nonnegative constraint in the
standard form $-\mathbf{b}\leq\mathbf{0}$ and introducing its Lagrange
multiplier $\boldsymbol{\lambda}\geq\mathbf{0}$ gives
\begin{equation}
	\label{eq:26}
	\mathcal{L}(\mathbf{b},\boldsymbol{\lambda})
	=
	f(\mathbf{b})
	-
	\boldsymbol{\lambda}^{\mathrm T}\mathbf{b}.
\end{equation}
At a KKT point
$(\widehat{\mathbf{b}},\widehat{\boldsymbol{\lambda}})$, stationarity
with respect to $\mathbf{b}$ requires
\begin{equation}
	\label{eq:stationarity}
	\nabla_{\mathbf{b}}
	\mathcal{L}
	(\widehat{\mathbf{b}},\widehat{\boldsymbol{\lambda}})
	=
	\mathbf{C}^{\mathrm T}
	\left(\mathbf{C}\widehat{\mathbf{b}}-\mathbf{y}\right)
	-
	\widehat{\boldsymbol{\lambda}}
	=
	\mathbf{0}.
\end{equation}
Together with primal
feasibility, dual feasibility, and complementary slackness, the NNLS
solution satisfies the Karush--Kuhn--Tucker conditions
\begin{equation}
	\label{eq:27}
	\widehat{\mathbf{b}}\geq\mathbf{0},~
	\widehat{\boldsymbol{\lambda}}
	=
		\nabla f(\mathbf{b})\geq\mathbf{0},~
	\widehat{\mathbf{b}}\odot	\nabla f(\mathbf{b})
	=
	\mathbf{0},
\end{equation}

The KKT conditions in \eqref{eq:27} characterize the NNLS optimum. To
obtain this optimum numerically, initialize the angular coefficient
vector as $\mathbf{b}^{(0)}=\mathbf{0}_{N\times1}$, which means that no
candidate angle is activated before the iteration. At the $k$-th
iteration, first evaluate the gradient at the current point and then
take a gradient step followed by a projection onto the nonnegative
orthant:
\begin{equation}
	\label{eq:28}
	\begin{aligned}
		\mathbf{b}^{(0)}
		&=
		\mathbf{0}_{N\times1},\\
		\mathbf{g}^{(k)}
		&=
		\nabla f\!\left(\mathbf{b}^{(k)}\right)
		=
		\mathbf{C}^{\mathrm T}
		\left(\mathbf{C}\mathbf{b}^{(k)}-\mathbf{y}\right),\\
		\mathbf{b}^{(k+1)}
		&=
		\left[
		\mathbf{b}^{(k)}
		-
		\mu\mathbf{C}^{\mathrm T}
		\left(\mathbf{C}\mathbf{b}^{(k)}-\mathbf{y}\right)
		\right]_{+},
	\end{aligned}
\end{equation}
Thus, the negative-gradient step
reduces the covariance-fitting objective, whereas the projection sets
every negative trial coefficient to zero and keeps
$\mathbf{b}^{(k+1)}\geq\mathbf{0}$.

To connect the iteration directly with the KKT conditions, define its
projected-gradient residual and stopping rule as
\begin{equation}
	\label{eq:29}
	\mathbf{b}^{(k)}
	-
		\left[
	\mathbf{b}^{(k)}
	-
	\mu\mathbf{C}^{\mathrm T}
	\left(\mathbf{C}\mathbf{b}^{(k)}-\mathbf{y}\right)
	\right]_{+}
	\leq
	\epsilon_{\mathrm{tol}}.
\end{equation}
 When the stopping rule in
\eqref{eq:29} is met, set $\widehat{\mathbf{b}}=\mathbf{b}^{(k)}$.

The converged NNLS estimate is continuous and is not necessarily binary
due to noise, finite covariance observations, and correlations between
adjacent dictionary columns. Since the number of propagation paths
$L_s$ is assumed to be known, the $L_s$ largest peaks of
$\widehat{\mathbf{b}}$ are selected as the AoA support.

\subsection{Path-Gain Estimation and LoS Identification}
\label{subsec:ls_los}

Let $\widehat{\theta}_l$, $l=1,\ldots,L_s$, denote the AoAs returned by
NNLS or one of the benchmark estimators.  These angles form an unordered
set: at this stage, the AoA estimator determines which angular responses
are present, but it neither recovers their complex path gains nor
identifies the LoS path.

The complex path gains are therefore recovered from the original pilot
observation in \eqref{eq:6}. Recall that $\mathbf{g}\in
\mathbb{C}^{M\times1}$ already denotes the channel vector over the $M$
activated ports. Right-multiplying \eqref{eq:6} by
$\mathbf{s}^{\mathrm H}\in\mathbb{C}^{L\times1}$ gives
\begin{align}
	\label{eq:pilot_despreading}
	\mathbf{Y}\mathbf{s}^{\mathrm H}
=
	\sqrt{\eta\beta}\,
	\mathbf{g}\mathbf{s}\mathbf{s}^{\mathrm H}
	+
	\mathbf{N}\mathbf{s}^{\mathrm H}
=
	\sqrt{\eta\beta}\,
	\mathbf{g}
	+
	\mathbf{N}\mathbf{s}^{\mathrm H},
\end{align}
Dividing \eqref{eq:pilot_despreading} by $\sqrt{\eta\beta}$ yields the
despread channel observation
\begin{align}
	\label{eq:despread_channel}
	\widetilde{\mathbf{g}}
	=
	\frac{\mathbf{Y}\mathbf{s}^{\mathrm H}}
	{\sqrt{\eta\beta}}
	=
	\mathbf{g}+\widetilde{\mathbf{n}},~
	\widetilde{\mathbf{n}}
	=
	\frac{\mathbf{N}\mathbf{s}^{\mathrm H}}
	{\sqrt{\eta\beta}},
\end{align}
where both $\widetilde{\mathbf{g}}$ and
$\widetilde{\mathbf{n}}$ belong to $\mathbb{C}^{M\times1}$.

For each detected angle, evaluate the steering vector defined in
\eqref{eq:12}. Collecting these steering vectors gives
\begin{equation}
	\label{eq:estimated_steering_matrix}
	\widehat{\mathbf{W}}
	=
	\begin{bmatrix}
		\mathbf{w}(\widehat{\theta}_1) &
		\mathbf{w}(\widehat{\theta}_2) &
		\cdots &
		\mathbf{w}(\widehat{\theta}_{L_s})
	\end{bmatrix}
	\in\mathbb{C}^{M\times L_s}.
\end{equation}
The ordering of the entries in $\mathbf{x}\in\mathbb{C}^{L_s\times1}$
is chosen to match the column ordering of
$\widehat{\mathbf{W}}$. This common permutation does not change the
channel representation in \eqref{eq:2}.

Using $\mathbf{g}=\mathbf{W}\mathbf{x}$ from \eqref{eq:2}, the
despread observation in \eqref{eq:despread_channel} can be written
without omitting the angle-estimation error as
\begin{align}
	\label{eq:path_gain_observation}
	\widetilde{\mathbf{g}}
	=
	\mathbf{W}\mathbf{x}+\widetilde{\mathbf{n}}
=
	\widehat{\mathbf{W}}\mathbf{x}
	+
	\boldsymbol{\varepsilon},~
	\boldsymbol{\varepsilon}
	\triangleq
	\left(\mathbf{W}-\widehat{\mathbf{W}}\right)\mathbf{x}
	+
	\widetilde{\mathbf{n}},
\end{align}
where $\boldsymbol{\varepsilon}\in\mathbb{C}^{M\times1}$ combines the
steering-matrix mismatch caused by imperfect AoA estimation and the
effective pilot noise. In the ideal on-grid and noiseless case,
$\widehat{\mathbf{W}}=\mathbf{W}$ and
$\boldsymbol{\varepsilon}=\mathbf{0}$.

With $\widetilde{\mathbf{g}}$ and $\widehat{\mathbf{W}}$ known, the
unknown complex path-gain vector is estimated by solving
\begin{equation}
	\label{eq:path_gain_ls}
	\widehat{\mathbf{x}}
	=
	\underset{\mathbf{x}\in\mathbb{C}^{L_s\times1}}{\arg\min}
	\left\|
	\widetilde{\mathbf{g}}
	-
	\widehat{\mathbf{W}}\mathbf{x}
	\right\|_2^2.
\end{equation}
Setting the derivative of the real-valued objective in
\eqref{eq:path_gain_ls} with respect to $\mathbf{x}$ to zero gives
\begin{equation}
	\label{eq:path_gain_normal_equation}
	\widehat{\mathbf{W}}^{\mathrm H}
	\left(
	\widehat{\mathbf{W}}\widehat{\mathbf{x}}
	-
	\widetilde{\mathbf{g}}
	\right)
	=
	\mathbf{0},
\end{equation}
the LS solution is:
\begin{align}
	\label{eq:path_gain_closed_form}
	\widehat{\mathbf{x}}
	&=
	\left(
	\widehat{\mathbf{W}}^{\mathrm H}\widehat{\mathbf{W}}
	\right)^{-1}
	\widehat{\mathbf{W}}^{\mathrm H}\widetilde{\mathbf{g}},
\end{align}
consequently, $\widehat{x}_l$ is the complex gain
associated with $\widehat{\theta}_l$, and
$|\widehat{x}_l|^2$ is the corresponding estimated path power.
\begin{figure}[!t]
	\centering
	\includegraphics[width=\columnwidth]{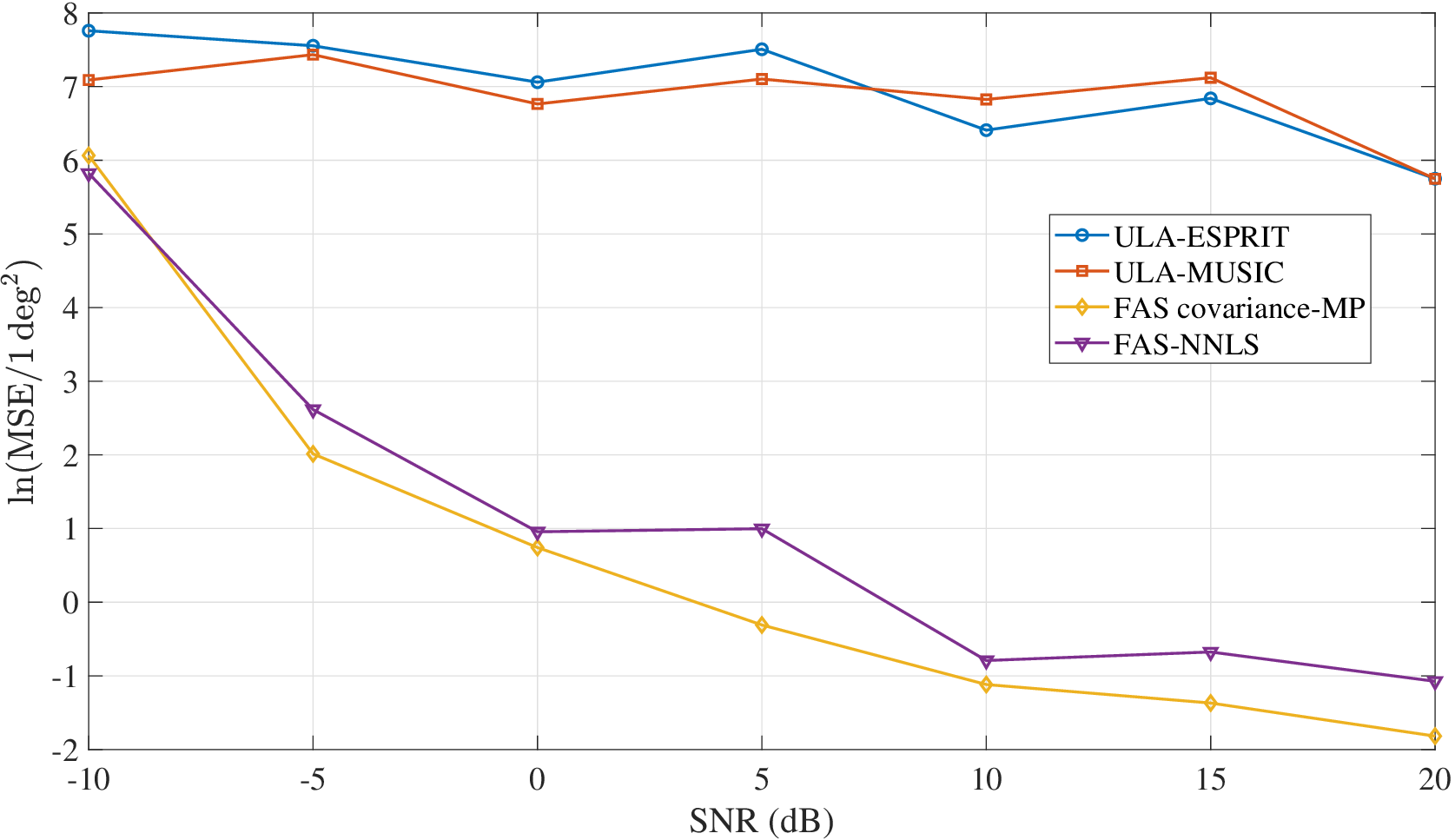}
	\caption{Natural logarithm of the mean squared AoA error.}
	\label{fig:single_path_mse}
\end{figure}

According to the known Rice factor $K$, each detected path is
successively treated as a candidate LoS path. The corresponding
estimated LoS-to-NLoS power ratio is defined as
\begin{equation}
	\widehat{K}_l
	=
	\frac{|\widehat{x}_l|^2}
	{\displaystyle
		\sum_{\substack{q=1\\q\neq l}}^{L_s}
		|\widehat{x}_q|^2},~ l=1,\ldots,L_s.
\end{equation}
The LoS path is identified as the candidate whose estimated power
ratio is closest to the known Rice factor, i.e.,
\begin{equation}
	\widehat{l}_{\mathrm{LoS}}
	=
	\underset{1\leq l\leq L_s}{\arg\min}
	\left|
	\widehat{K}_l-K
	\right|.
\end{equation}
All remaining detected paths are classified as NLoS paths.

\section{Simulation Results}
\label{sec:simulation_results}
\begin{figure}[!t]
	\centering
	\includegraphics[width=\columnwidth]{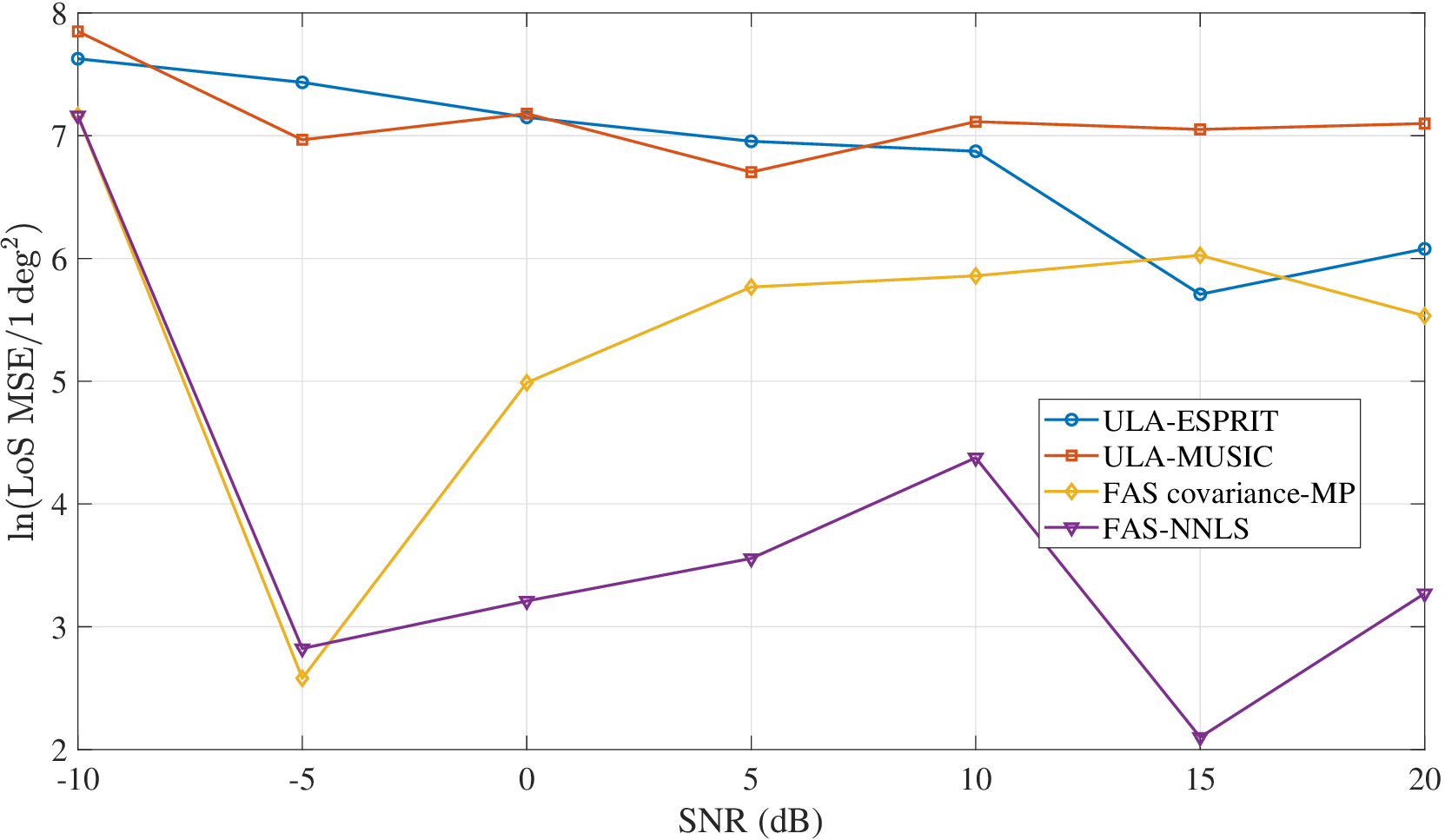}
	\caption{Natural logarithm of the LoS mean squared AoA error.}
	\label{fig:two_path_los_mse}
\end{figure}

\begin{figure}[!t]
	\centering
	\includegraphics[width=\columnwidth]{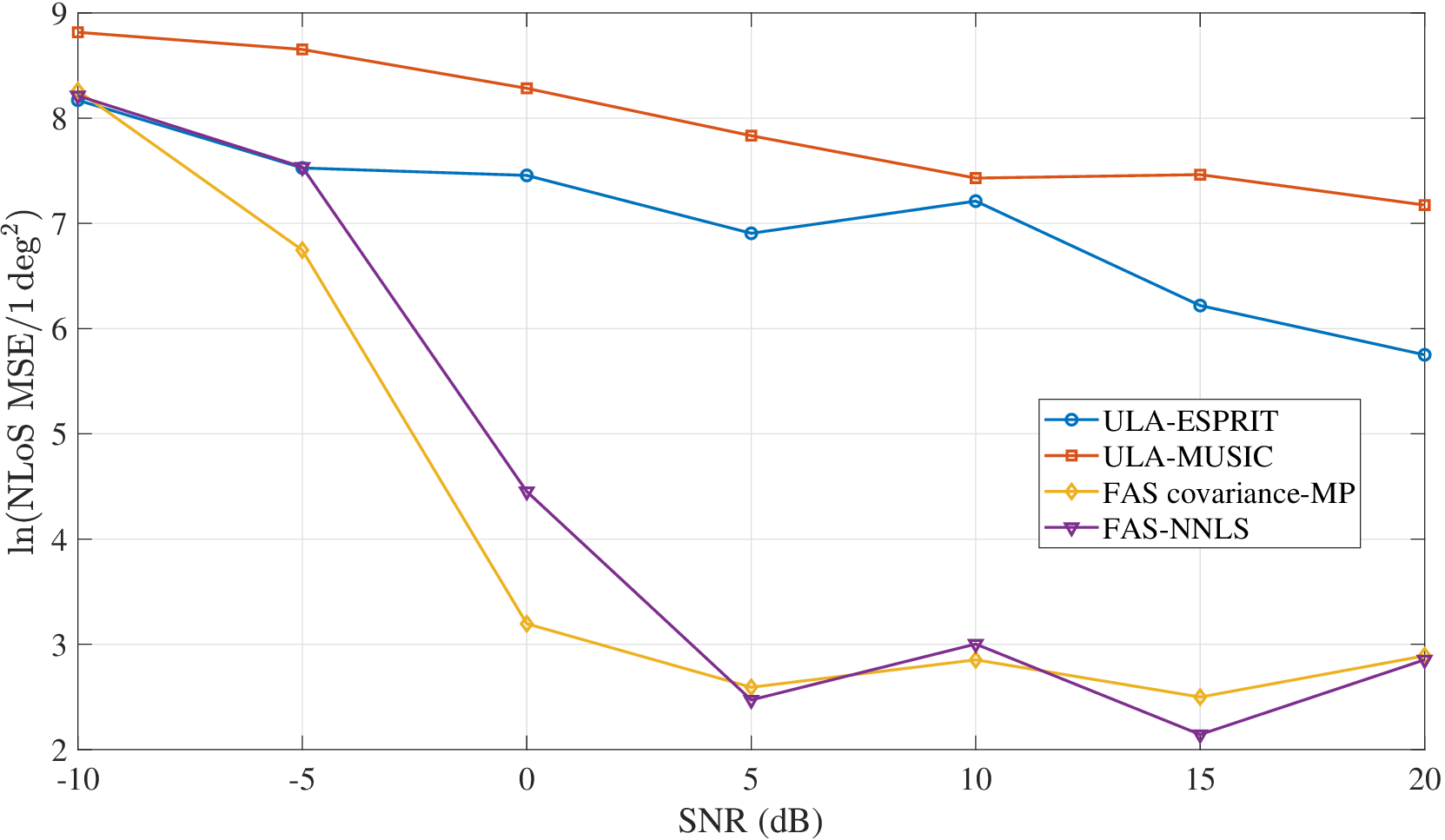}
	\caption{Natural logarithm of the NLoS mean squared AoA error.}
	\label{fig:two_path_nlos_mse}
\end{figure}
The proposed FAS-NNLS estimator is compared with ULA-ESPRIT,
ULA-MUSIC, and FAS covariance-MP introduced in
Subsection~\ref{subsec:benchmark_algorithms}. In all simulations,
$M=5$ antennas or activated ports and $L=100$ snapshots are used. The
ULA has half-wavelength spacing, while the FAS contains $N_f=10$ ports
over $W=2$ wavelengths and activates
$\mathcal{P}=\{0,1,4,7,9\}$. The candidate-angle grid spans
$[-90^{\circ},90^{\circ}]$ with a spacing of $0.1^{\circ}$, and the
total average path power is normalized to one. For each SNR, the
results are averaged over $N_{\mathrm{MC}}=100$ independent Monte
Carlo trials with randomly generated AoAs and noise. For both channel
settings, the SNR ranges from $-10$ to $20$ dB in $5$-dB increments,
while the two-path case consists of one LoS and one NLoS path with
$K=2$. After the LoS/NLoS identification in
Subsection~\ref{subsec:ls_los}, the comparison metric for path type
$q$ is the unnormalized mean squared AoA error
\begin{equation}
	\label{eq:aoa_mse_metric}
	\operatorname{MSE}_{q}(\rho)
	=
	\frac{1}{N_{\mathrm{MC}}}
	\sum_{i=1}^{N_{\mathrm{MC}}}
	\left(
		\widehat{\theta}_{q,i}(\rho)-\theta_{q,i}
	\right)^2,
\end{equation}
where $\rho$ denotes the SNR and the error is measured in
$\mathrm{deg}^{2}$. For visualization, each curve reports
$\ln\!\left(\mathrm{MSE}_{q}(\rho)/(1\,\mathrm{deg}^{2})\right)$.

For the single-path case, Fig.~\ref{fig:single_path_mse} shows that both
FAS covariance estimators exhibit comparable performance and consistently
outperform the ULA subspace methods throughout the investigated SNR range.
As the SNR increases, both FAS methods reduce the logarithmic MSE below
zero, whereas ULA-ESPRIT and ULA-MUSIC retain substantially larger
residual errors.

For the channel containing one LoS and one NLoS path,
Figs.~\ref{fig:two_path_los_mse} and
\ref{fig:two_path_nlos_mse} report the two path errors separately. In
the moderate-to-high-SNR region, the NLoS estimation error of FAS-NNLS
is close to that of FAS covariance-MP and remains markedly below those
of the ULA benchmarks. More importantly, FAS-NNLS provides a clear
improvement over FAS covariance-MP for the LoS path across most of the
tested SNR range, with its LoS error remaining close to zero once the
noise level is sufficiently reduced.

Overall, the nonuniform FAS produces a difference coarray with a larger
effective aperture and more spatial degrees of freedom than the
conventional $M$-element ULA. By combining this virtual aperture with
the nonnegative covariance model, NNLS extracts reliable angular
support, improves the subsequent LS path-gain estimation, and thereby
enables favorable full FAS channel-reconstruction performance.

\section{Conclusion}
\label{sec:conclusion}

This paper addressed full-aperture FAS channel reconstruction from pilot
observations at a subset of nonuniformly activated ports. Vectorizing the
received covariance mapped the selected ports to a virtual difference
coarray, and the angular-support estimation problem was cast as an NNLS
problem over nonnegative combinations of virtual steering responses. Numerical results demonstrate that the virtual aperture and spatial degrees of
freedom afforded by the FAS difference coarray translate directly into more
accurate angular-support recovery and more reliable parametric channel
reconstruction.

\balance
\bibliographystyle{IEEEtran}
\bibliography{yinyong}

@article{wong2022extra,
  title={Extra-large {MIMO} enabling slow fluid antenna massive access for millimeter-wave bands},
  author={Wong, Kai-Kit and Tong, Kin-Fai and Chen, Yu and Zhang, Yangyang},
  journal={Electronics Letters},
  volume={58},
  number={25},
  pages={1016--1018},
  year={2022},
}

@article{zheng2023fast,
  author={Zheng, Chundi and Yu, Meiyi and Shan, Jiaolong and Wang, Aiguo and Chen, Huihui},
  title={Fast Sparse Non-Negative Least Squares via {ADMM} for High Resolution {DOA} Estimation},
  journal={IEEE Sensors Journal},
  volume={23},
  number={4},
  pages={3901--3910},
  year={2023},
}

@ARTICLE{11155198,
  author={Zhang, Zhentian and Wong, Kai-Kit and Dang, Jian and Zhang, Zaichen and Chae, Chan-Byoung},
  journal={IEEE Journal on Selected Areas in Communications}, 
  title={On Fundamental Limits for Fluid Antenna-Assisted Integrated Sensing and Communications for Unsourced Random Access}, 
  year={2026},
  volume={44},
  number={},
  pages={136-149},
}

@article{ZZT_SlowFluidURA_2025,
	author  = {Z. Zhang and K.-K. Wong and J. Dang and Z. Zhang and C. Masouros and C.-B. Chae},
	title   = {On Fundamental Limits of Slow-Fluid Antenna Multiple Access for Unsourced Random Access},
	journal = {IEEE Wireless Commun. Lett.},
	volume  = {14},
	number  = {11},
	pages   = {3455--3459},
	year    = {2025}
}

@article{schmidt1986music,
  author={Schmidt, Ralph O.},
  title={Multiple Emitter Location and Signal Parameter Estimation},
  journal={IEEE Transactions on Antennas and Propagation},
  volume={34},
  number={3},
  pages={276--280},
  year={1986},
}

@article{roy1989esprit,
  author={Roy, Richard and Kailath, Thomas},
  title={{ESPRIT}---Estimation of Signal Parameters via Rotational Invariance Techniques},
  journal={IEEE Transactions on Acoustics, Speech, and Signal Processing},
  volume={37},
  number={7},
  pages={984--995},
  year={1989},
}

@ARTICLE{9449971,
  author={Leite, Wesley S. and Lamare, Rodrigo C. de},
  journal={IEEE Transactions on Aerospace and Electronic Systems}, 
  title={List-Based OMP and an Enhanced Model for DOA Estimation With Nonuniform Arrays}, 
  year={2021},
  volume={57},
  number={6},
  pages={4457-4464},
}

@article{wong2023preliminaries,
  author={Wong, Kai-Kit and New, Wee Kiat and Xu, Hao and Tong, Kin-Fai and Chae, Chan-Byoung},
  title={Fluid Antenna System---Part {I}: Preliminaries},
  journal={IEEE Communications Letters},
  volume={27},
  number={8},
  pages={1919--1923},
  year={2023},
}

@article{xu2024l3scr,
  author={Xu, Hao and Zhou, Gui and Wong, Kai-Kit and New, Wee Kiat and Wang, Chao and Chae, Chan-Byoung and Murch, Ross and Jin, Shi and Zhang, Yangyang},
  title={Channel Estimation for {FAS}-Assisted Multiuser {mmWave} Systems},
  journal={IEEE Communications Letters},
  volume={28},
  number={3},
  pages={632--636},
  year={2024},
}

@article{new2025oversampling,
  author={New, Wee Kiat and Wong, Kai-Kit and Xu, Hao and Ghadi, Farshad Rostami and Murch, Ross and Chae, Chan-Byoung},
  title={Channel Estimation and Reconstruction in Fluid Antenna System: Oversampling Is Essential},
  journal={IEEE Transactions on Wireless Communications},
  volume={24},
  number={1},
  pages={309--322},
  year={2025},
}

@article{ZZT_JointActivityChannel_2026,
	author  = {Z. Zhang and J. Dang and D. Morales-Jimenez and H. Jiang and Z. Zhang and C. Masouros and C.-B. Chae},
	title   = {Joint Activity Detection and Channel Estimation for Fluid Antenna System Exploiting Geographical and Angular Information},
	journal = {IEEE J. Sel. Topics Signal Process.},
	volume  = {20},
	number  = {3},
	pages   = {354--370},
	year    = {2026}
}

@article{ZZT_FiniteBlocklengthFAS_2026,
	author  = {Z. Zhang and K.-K. Wong and D. Morales-Jimenez and H. Jiang and H. Xu and C. Masouros and Z. Zhang},
	title   = {Finite-Blocklength Fluid Antenna Systems},
	journal = {IEEE Trans. Wireless Commun.},
	year    = {2026},
	note    = {Early Access},
	doi     = {10.1109/TWC.2026.3723456}
}

@article{ZZT_FiniteBlocklengthCorrelation_2026,
	author  = {Z. Zhang and K.-K. Wong and D. Morales-Jimenez and H. Jiang and P. Ramirez-Espinosa and C.-B. Chae and C. Masouros},
	title   = {Finite-Blocklength Fluid Antenna Systems with Spatial Block-Correlation Channel Model},
	journal = {IEEE Wireless Commun. Lett.},
	volume  = {15},
	pages   = {1911--1915},
	year    = {2026}
}

@ARTICLE{fas-twc-21,
	author  = {K. K. Wong and A. Shojaeifard and K.-F. Tong and Y. Zhang},
	title   = {Fluid Antenna Systems},
	journal = {IEEE Trans. Wireless Commun.},
	volume  = {20},
	number  = {3},
	pages   = {1950--1962},
	month   = mar,
	year    = {2021}
}

@ARTICLE{kit_electronic,
	author  = {K. K. Wong and K. F. Tong and Y. Chen and Y. Zhang},
	title   = {Closed-Form Expressions for Spatial Correlation Parameters for Performance Analysis of Fluid Antenna Systems},
	journal = {Electron. Lett.},
	volume  = {58},
	number  = {11},
	pages   = {454--457},
	month   = apr,
	year    = {2022}
}

@article{ZZT_FiniteAperturePlanar_2026,
	author  = {Z. Zhang and J. Xu and K.-K. Wong and H. Jiang and Z. Zhang and H. Shin},
	title   = {Finite-Aperture Planar Fluid Antenna Array},
	journal = {arXiv preprint},
	year    = {2026},
	url     = {https://arxiv.org/abs/2605.22040}
}

@article{ZZT_FiniteApertureDesign_2026,
	author  = {Z. Zhang and K.-K. Wong and H. Jiang and F. Rostami Ghadi and H. Shin and Y. Zhang},
	title   = {Finite-Aperture Fluid Antenna Array Design: Analysis and Algorithm},
	journal = {IEEE Wireless Commun. Lett.},
	volume  = {15},
	pages   = {3199--3203},
	year    = {2026}
}

@article{ZZT_LearnedAMP_2026,
	author  = {Y. Wu and Z. Zhang and H. Jiang and K.-K. Wong and C.-B. Chae},
	title   = {Learned-Approximate Message Passing Under Karhunen--Lo{\`e}ve Modeling for Fluid Antenna Systems},
	journal = {IEEE Wireless Commun. Lett.},
	volume  = {15},
	pages   = {2719--2723},
	year    = {2026}
}

@inproceedings{FAA_XJY,
	author    = {J. Xu and Z. Zhang and J. Dang and H. Jiang and Z. Zhang},
	title     = {Fluid Antenna-Enhanced Flexible Beamforming},
	booktitle = {Proc. IEEE Wireless Commun. Netw. Conf. Workshops (WCNCW)},
	year      = {2026},
	pages     = {1--6}
}

@inproceedings{ML_WCNC,
	author    = {H. Liang and Z. Zhang and J. Dang and H. Jiang and Z. Zhang},
	title     = {Neural Networks-Enabled Channel Reconstruction for Fluid Antenna Systems: A Data-Driven Approach},
	booktitle = {Proc. IEEE Wireless Commun. Netw. Conf. Workshops (WCNCW)},
	year      = {2026},
	pages     = {1--6}
}

@INPROCEEDINGS{11586648,
	author={Liang, Haoyu and Zhang, Zhentian and Dang, Jian and Jiang, Hao and Zhang, Zaichen},
	booktitle={2026 IEEE International Conference on Communications Workshops (ICC Workshops)}, 
	title={DoA Estimation Based on Deep Learning-MUSIC Hybrid Algorithm with Fluid Antenna for Closely Spaced Signals}, 
	year={2026},
	volume={},
	number={},
	pages={1-6},
	doi={10.1109/ICCWorkshops63917.2026.11586648}}

@ARTICLE{11556494,
	author={Chen, Sen and Zhang, Zhentian and Jiang, Hao and Wong, Kai-Kit and Li, An and Rostami Ghadi, Farshad and Shin, Hyundong},
	journal={IEEE Commun. Lett.}, 
	title={Near-Field Beamforming and Port Selection for Fluid Antenna Systems}, 
	year={2026},
	volume={30},
	number={},
	pages={2248-2252},
	doi={10.1109/LCOMM.2026.3702003}}

@article{ZZT_GeometryReconstruction_2026,
	author  = {Z. Zhang and K.-K. Wong and K. Meng and D. Morales-Jimenez and H. Jiang and C. Masouros and H. Shin and Z. Zhang},
	title   = {Geometry-Structured Channel Reconstruction for Conventional and Fluid Antenna Systems: Bayesian Inference and Fundamental Limits},
	journal = {arXiv preprint},
	year    = {2026},
	url     = {https://arxiv.org/abs/2606.04001}
}

@article{ZZT_CRBActivityDetection_2026,
	author  = {Z. Zhang and K.-K. Wong and H. Jiang and C. Masouros and C.-B. Chae},
	title   = {Cram{\'e}r--Rao Bounds for Activity Detection in Conventional and Fluid Antenna Systems},
	journal = {IEEE Wireless Commun. Lett.},
	volume  = {15},
	pages   = {3059--3063},
	year    = {2026}
}

@article{ZZT_DualSideFAS_2026,
	author  = {Z. Zhang and Y. Wu and K.-K. Wong and H. Jiang and A. Li},
	title   = {Jointly Correlated Dual-Side Fluid Antenna System},
	journal = {IEEE Wireless Commun. Lett.},
	volume  = {15},
	pages   = {4370--4374},
	year    = {2026}
}

\end{document}